\documentclass{jfm}
\usepackage{graphicx}
\usepackage{epstopdf, epsfig}
\usepackage{soul}
\usepackage{siunitx}
\usepackage{array}
\usepackage{subfig}
\usepackage{graphicx}
\usepackage{natbib}
\usepackage{hyperref}
\usepackage{bm}
\usepackage{booktabs}
\usepackage{xcolor}
\usepackage{moreverb}
\setcitestyle{numbers,open={[},close={]}}
\usepackage{amssymb}
\usepackage{amsmath}
\usepackage{mathtools}
\usepackage{multirow}
\usepackage{array}
\usepackage{wasysym}
\usepackage{booktabs}
\usepackage{gensymb}
\usepackage{caption}
\usepackage{graphicx}
\usepackage{tabularx}
\usepackage{stackengine}
\usepackage{pbox}

\definecolor{applegreen}{rgb}{0.55, 0.71, 0.0}

\shorttitle{Spatiotemporal Instabilities in Discontinuous Shear Thickening}
\shortauthor{Angerman et al.}

\title{Numerical Simulations of Spatiotemporal Instabilities in Discontinuous Shear Thickening Fluids}

\author{Peter Angerman\aff{1,2}
  \corresp{\email{2037484@swansea.ac.uk}},
  Bjornar Sandnes\aff{1}
  Ryohei Seto\aff{3,4,5}
 \and Marco Ellero\aff{2,6,7}}

\affiliation{\aff{1}Complex Fluids Research Group, Department of Chemical Engineering, Swansea University, Swansea SA1 8EN, United Kingdom
\aff{2}Basque Center for Applied Mathematics (BCAM), Alameda de Mazarredo 14, 48009 Bilbao, Spain
\aff{3}Wenzhou Key Laboratory of Biomaterials and Engineering, Wenzhou Institute, University of Chinese Academy of Sciences, Wenzhou, 325000, China
\aff{4}Oujiang Laboratory (Zhejiang Lab for Regenerative Medicine, Vision and Brain Health), Wenzhou, 325000, China
\aff{5}Graduate School of Information Science, University of Hyogo, 
Kobe, 650-0047, Japan
\aff{6}IKERBASQUE, Basque Foundation for Science, Calle de Mar\'{i}a D\'{i}az de Haro 3, 48013 Bilbao, Spain
\aff{7}Zienkiewicz Centre for Computational Engineering (ZCCE), Swansea University, Bay Campus, Swansea SA1 8EN, United Kingdom}

\begin{document}

\maketitle

\begin{abstract}
Discontinuous Shear Thickening (DST) fluids exhibit unique instability properties in a wide range of flow conditions. We present numerical simulations of a scalar model for DST fluids in a planar simple shear using the Smoothed Particle Hydrodynamics (SPH) approach. The model reproduces the spatially homogeneous instability mechanism based on the competition between the inertial and microstructural timescales, with good congruence to the theoretical predictions. Spatial inhomogeneities arising from a stress-splitting instability are rationalised within the context of local components of the microstructure evolution. Using this effect, the addition of non-locality in the model is found to produce an alternative mechanism of temporal instabilities, driven by the inhomogeneous pattern formation. The reported arrangement of the microstructure is generally in agreement with the experimental data on gradient pattern formation in DST. Simulations in a parameter space representative of realistic DST materials resulted in aperiodic oscillations in measured shear rate and stress, driven by formation of gap-spanning frictional structures.
\end{abstract}


\section{Introduction}

Discontinuous shear-thickening (DST) dense suspensions of non-Brownian particles are characterised by a rapid increase in viscosity, attributed to a transition from a frictionless state to a frictional state facilitated by a shear stress sufficient to overcome the characteristic stress scale $\sigma^\ast$, inherent to the particular nature of the suspension particles and solvent \citep{WC2014,Seto2013,Seto2014,Coling2017}.
The distinguishing feature of DST as opposed to continuous shear-thickening (CST) is the presence of a negative gradient of the flow curve ($d\dot{\gamma}/d\sigma<0$), due to the increase in viscosity spanning multiple orders of magnitude \citep{Jaeger2019}.
Recently, the stability properties of DST materials have been a subject of extensive research \citep{Morris2020}, partly because of their potential for understanding the role of internal microstructure in flows applicable to rheometric conditions and large-scale free surface flows.

Instabilities in rheometric flows are of particular interest due to their role in the difficulties associated with the characterisation of DST materials \citep{Jaeger2019} under typical simple shear conditions.
In stress-controlled conditions, DST fluids exhibit aperiodic or almost periodic oscillations in the measured shear rate and normal stress \citep{Hermes2016}---a phenomenon often referred to as `rheochaos'.
The onset of the instability occurs at a critical stress and grows in strength up to a point, after which further increases in stress reduce the amplitude.


A plausible explanation for the observed temporal instability is the existence of an accompanying spatial instability. 
Dense frictional suspensions are understood to form contact force networks \citep{kcore,Seto2013}, facilitating stress transmission.
As such, it is reasonable to expect that large magnitude temporal fluctuations in normal stress are associated with formation, destruction, and rearrangement of the internal micro-structure.
Based on the multivariate nature of the flow curve, DST materials could be expected to undergo formation of bands along the vorticity direction \citep{Olmsted2008}; 
however, the existence of stable vorticity bands can be ruled out on the basis of particle pressure balance at the band interface \citep{Hermes2016}. 
In fact, instead of stable vorticity bands, transient vorticity banding has been reported \citep{Herle}. 


A further connection between spatial and temporal instabilities comes from recent literature utilising direct online measurements of the internal fluid state.
Force and pressure sensors have been used to detect localised regions of high and low normal pressure travelling in the direction of flow \citep{Nakanishi2016}. 
Ultrasonic echography has yielded direct imaging of bands propagating in the vorticity direction during temporal oscillations \citep{Manneville2018}. 
Perhaps the most illuminating direct imaging comes from recent work \citep{Vikram2017,Vikram2020,Vikram2022} using boundary stress microscopy (BSM), in which the stress exerted on the bottom plate in a parallel plate setup can be imaged in real time along with the movement of individual particles via particle image velocimetry (PIV). 
BSM imaging revealed regions of localised high stress (a peak magnitude of an order larger than the bulk) in simple and oscillatory shear. 
Tracking these solid-like phases (SLPs) over time allowed for inference of the existence of SLPs at the upper and lower plate boundaries moving with the velocity close to their proximate solid boundaries.


Continuum modelling of DST fluids ubiquitously begins with the definition of the characteristic S-shaped flow curve, primarily through the Wyart--Cates (WC) model \citep{WC2014}, with some exceptions \citep{Nakanishi,Bossis2017}. 
The WC model is a phenomenological approach, with solid microscopic evidence \citep{kcore}, that proposes that particles are held apart by a characteristic repulsive force $F_0$ which sets the characteristic stress scale $\sigma^\ast$. 
Below the characteristic stress scale, the particles are held apart in a low-viscosity, frictionless state. 
When a stress is applied above the onset stress, the particles enter frictional contact and rapidly increase the viscosity. 
The key parameter in the WC model, $f(\sigma)$, represents the fraction of particles in a frictional contact at a given stress. 
Most works on continuum modelling use this scalar parameter \citep{Nakanishi2016,
Kamrin2019,Fielding2018,
Darbois-Texier2020}, along with a microstructural evolution equation, which seeks to describe the dynamics of the frictionless-to-frictional transition.
Typically, the rate of evolution is a rate-dependent approach of $f$ towards the WC prescribed steady-state value $\hat{f}(\sigma)$ at a given stress.

One potential mechanism of temporal instabilities identified in the literature comes from competition between the timescales of system-inertia ($t_{\mathrm{i}}$) and microstructure formation ($t_{\mathrm{f}}$) \citep{Richards}. 
Stability analysis of the coupled dynamics of a stress-controlled plate and the commonly used local form of the microstructure evolution equation reveal that such a system can undergo a Hopf bifurcation. 
When the rate of formation is much higher than the rate of boundary deacceleration ($t_{\mathrm{i}} \gg t_{\mathrm{f}}$), an overshoot in the shear rate will induce rapid shear thickening, which in turn will further increase the material stress since the plate cannot decelerate sufficiently quickly, resulting in a feedback loop. 
In addition to the required ratio of timescales, negative gradients of the flow curve are also necessary, indicating that this mechanism is unique to DST materials. 
The primary limitation in the analysis associated with this mechanism is the inherent assumption of homogeneous space, which clearly does not account for the spatial pattern formation observed in real systems.
Furthermore, because of the two-dimensional aspect of the system in phase space, only periodic oscillatory solutions are possible, which limits the applicability to a narrow region in which periodicity is observed in experiments.


Alternatively, coupling of the simple evolution of the scalar microstructure with the dynamic migration of particles through the suspension balance model \citep{Fielding2018} has been shown to produce transient vorticity banding, uniquely for DST conditions. 
The observed pattern formation was shown to be in a good agreement with direct particle simulations; however, the continuum modelling was limited to a gradient--vorticity space, with assumption of spatial homogeneity in both flow and gradient directions.


In this work, we present a systematic study of the possible interactions between the microstructural, inertial, and stress-diffusive timescales, by framing pattern formation in context of a competition between local and non-local contributions to the microstructure evolution, in presence of DST rheology. 
In Sections \ref{cont model}, we outline the continuum model and numerical implementation, respectively. 
In Section \ref{dominant non-local}, we consider the case of dominant non-locality for the purpose of validation of our SPH model against a temporal instability mechanism proposed in the literature. 
In Section \ref{weak non-local} we demonstrate the stress-splitting instability inherent in the topology of DST flow curves.
Finally, in Section \ref{moderate non-local} we consider regimes of mixed non-locality where spatial pattern formation is possible and present an alternative mode of temporal instability, driven by variations in microstructural spatial patterns within the geometry.

 \section{Method}
 
 \subsection{DST - continuum scalar model}
 
 \label{cont model}
 The governing equations, written in the Lagrangian frame, are 
 \begin{gather}
 \frac{D\rho}{Dt}=-\rho\nabla \cdot \boldsymbol{v}, \\
\frac{D\boldsymbol{v}}{Dt}=\frac{1}{\rho}\nabla\cdot\boldsymbol{\sigma}+\boldsymbol{F}, 
\label{momentum_equation}
\\ 
 \boldsymbol{\sigma}=-p\boldsymbol{I}+
 \eta \left(\nabla \boldsymbol{v}+\nabla\boldsymbol{v}^{\mathsf{T}}\right),
\end{gather}
where $D/Dt \equiv \partial/\partial t + \boldsymbol{v}\cdot\nabla$ is the material derivative,  $\rho(\boldsymbol{x},t)$ is the density field, $\boldsymbol{v}(\boldsymbol{x},t)$ is the velocity field, $\boldsymbol{\sigma}(\boldsymbol{x},t)$ is the stress field, $p$ is the pressure, and $\boldsymbol{F}$ is a body force acting on the fluid.


The viscosity of the suspension is defined by the Maron--Pierce relation \citep{Maron}
\begin{equation}
\label{Maron–Pierce}
\eta(\phi,\phi^{\mathrm{J}},t)
=
\eta_{\mathrm{s}}
  \left(1-\frac{\phi}{\phi^{\mathrm{J}} } \right)^{-2},
\end{equation}
where $\eta(\phi,\phi^{\mathrm{J}},t)$ and $\eta_{\mathrm{s}} $ are, respectively, the suspension and constant solvent viscosities. 
The divergence volume fraction, $\phi^{\mathrm{J}}$, is used to introduce shear thickening by interpolating its value between the frictionless state ($\phi^0$) and the frictional state ($\phi^{\mathrm{m}}$),
\begin{equation}
\label{jamming_point}
 \phi^{\mathrm{J}} (f(\boldsymbol{x},t))=\phi^{\mathrm{m}} 
 f(\boldsymbol{x},t)+\phi^0
 \{ 1-f(\boldsymbol{x},t) \}.
\end{equation}
Here, the interpolating state parameter, $f(\boldsymbol{x},t)$, is interpreted as the fraction of particles in frictional contact at a given position in space, as defined in the WC model. 
The dynamics of the state parameter is defined via the microstructure evolution equation:
\begin{gather}
    \frac{Df (\boldsymbol{x},t)}{Dt}=k_{\mathrm{f}} 
    \dot{\gamma}
    \left(\hat{f}(\boldsymbol{x},t)
    -f(\boldsymbol{x},t)\right)+\alpha\nabla^2 f(\boldsymbol{x},t), 
    \label{eq: f evolution}\\
    \hat{f}(\boldsymbol{x},t)=
    \exp\biggl(
    -\left[\frac{\sigma^{\ast}}{\sigma(\boldsymbol{x},t)}\right]^2 \biggr),
    \label{WC eq}
\end{gather}
which consists of a local term modelling the approach of the system towards a stress-prescribed WC steady state, $\hat{f}(\boldsymbol{x},t)$, and a non-local term modelling stress-diffusive components of microstructure formation in space. 
The local contributions are tuned with a microstructural rate constant, $k_{\mathrm{f}}$, whereas the non-local dynamics are set by the microstructure diffusivity parameter, $\alpha$. 
The fluid stress is calculated as $\sigma=\eta\dot{\gamma}$, where $\dot{\gamma}(\boldsymbol{x},t)
=\sqrt{\dot{\boldsymbol{\gamma}}:\dot{\boldsymbol{\gamma}}/2}$ is the local shear rate calculated as the second invariant of the strain rate tensor 
$\dot{\boldsymbol{\gamma}}(\boldsymbol{x},t)=\left(\nabla \boldsymbol{v}+\nabla\boldsymbol{v}^{\mathsf{T}}\right)/2$. 
\subsection{Numerical Implementation}

In all simulations, unless otherwise noted, we use the WC parameters $\phi=0.54$, $\phi^{\mathrm{J}}=0.693$ and $\phi_{\mathrm{m}}=0.562$, resulting in a well-defined 'S'-shaped flow curve \citep{Angerman2024}. 
In Sections \ref{dominant non-local} and \ref{weak non-local}, we simulate a square box with a gap height of $h=\SI{0.01}{\metre}$, while, in Section \ref{moderate non-local}, we use a wide box $l = 3 h$
with the same gap height to fully capture the resulting spatial pattern. 
In all simulations, the numerical resolution is kept constant at $\Delta x=h/50$.

We simulate simple shear in stress-controlled mode by setting the velocity on the upper wall ($v_{\mathrm{w}}$) and evolving it over time:
\begin{equation}
    \frac{dv_{\mathrm{w}}}{dt}=k_{\mathrm{p}}
    (\sigma_{\mathrm{E}}-\sigma),
    \label{stress control}
\end{equation}
where $\sigma_{\mathrm{E}}$ is the target stress set-point, $\sigma$ is the shear stress measure at the upper wall, and $k_{\mathrm{p}}$ is a tuning parameter in the control scheme. A small value for $k_{\mathrm{p}}$ corresponds to slow adjustment of velocity in response to rheological changes (high system inertia). A large value for $k_{\mathrm{p}}$ corresponds to fast adjustment (low system inertia).
The microstructural field, $f(\boldsymbol{x})$, and the shear rate, $\dot{\gamma}$, are initialised, in most cases, from a steady state exactly on the WC flow curve corresponding to the stress set-point, with $f(\boldsymbol{x})$ uniform.

\subsection{Dimensional Analysis}
The dynamics of our model is captured by three characteristic timescales:
\begin{gather}
    t_{\mathrm{i}}=
    \frac{h}{k_{\mathrm{p}}\eta_{\mathrm{s}}}, 
    \qquad 
    t_{\mathrm{f}}
    =
    \frac{\eta_\mathrm{s}}{k_\mathrm{f} \sigma^\ast}, 
    \qquad 
    t_\mathrm{d}=\frac{L^2}{\alpha},
\end{gather}
where $t_\mathrm{i}$ is the system-inertia timescale, $t_\mathrm{f}$ represents the local microstructure evolution timescale, and $t_\mathrm{d}$ is the microstructure diffusive (or non-local) timescale. 
The relevant length scale, $L$, is the numerical resolution length, $\Delta x$, allowing us to interpret the non-local term as an artificial diffusive term, where the resolution can be scaled out as $\alpha=\alpha' L^2$. 
We proceed by defining the dimensionless parameters
\begin{gather}
    \Sigma=\frac{\sigma}{\sigma^\ast}, \qquad 
    \dot{\Gamma}=\frac{\dot{\gamma}\eta_{\mathrm{s}}}{\sigma^\ast}, \qquad 
    \epsilon=\frac{t_{\mathrm{f}}}{t_{\mathrm{i}}}, \qquad 
    \omega=\frac{t_{\mathrm{d}}}{t_{\mathrm{i}}},
\end{gather}
where $\Sigma$ is the dimensionless stress, $\dot{\Gamma}$ is the dimensionless shear rate, $\epsilon$ and $\omega$ are local and non-local microstructural timescales, respectively, normalised with the inertial timescale.

The corresponding nondimensional forms of the microstructural evolution equation \eqref{eq: f evolution} and the stress control equation \eqref{stress control} are given by:
\begin{gather}
     \frac{df (\boldsymbol{\Tilde{x}},t)}{d\tilde{\tau}}=\frac{\dot{\Gamma}}{\epsilon}
    \left(\hat{f}(\boldsymbol{\tilde{x}},t)
    -f(\boldsymbol{\tilde{x}},t)\right)+\frac{1}{\omega}\nabla^2f(\boldsymbol{\tilde{x}},t), 
    \label{eq: fdim evolution}\\
    \frac{d\dot{\Gamma}}{d\tilde{\tau}}=\Sigma_{\mathrm{E}}-\Sigma,
    \label{ss dim}
\end{gather}
where $\tilde{\boldsymbol{x}}=\boldsymbol{x}/L$ and $\tilde{\tau}=t/t_{\mathrm{i}}$.

\section{Numerical results}
\subsection{Dominant Non-Locality}
\label{dominant non-local}
To classify relative magnitudes of local and non-local contributions to the evolution of microstructure, we define a characteristic stress-diffusive length scale, 
\begin{equation}
    l_{\mathrm{c}}=\sqrt{\frac{\alpha}{k_{\mathrm{f}} \dot{\gamma}}},
    \label{lc}
\end{equation}
which determines the spatial scale of potential pattern formation. 
In this section, we consider an entirely homogeneous state by setting $l_{\mathrm{c}} \approx h$. 
We set $k_{\mathrm{f}} = 1$, and since in the flow curve range of interest $\dot{\gamma}=O(10^{-1})$, using $\alpha=10^{-5}$ allows for characteristic length scale spanning approximately the entire spatial domain.


In the extreme case of dominant non-locality, we can expect an entirely homogeneous state, and due to the vanishing gradients of $f(\boldsymbol{x})$, we can simplify the dynamic model equations by neglecting the effects of the stress-diffusive term (last term in Eq. (\ref{eq: fdim evolution})) on the stability of the dynamics. 
Doing so simplifies the equations to a system directly equivalent to the one studied by \citet{Richards}, allowing a comparison to their stability analysis. 
Following their analysis, the stability criterion for the mechanism based on the competition of the system-inertia timescale $(t_{\mathrm{i}})$ and the microstructural formation timescale $(t_{\mathrm{f}})$ can be written as
\begin{equation}
 \frac{t_{\mathrm{f}}}{t_{\mathrm{i}} } < -\dot{\Gamma}\frac{d\dot{\Gamma}}{d\Sigma_{\mathrm{E}}},
\end{equation}
which, if satisfied, indicates unstable dynamics and the existence of periodic solutions. 
The condition can be rewritten into a more convenient formulation with the overall stability parameter
\begin{equation}
    \mu=-\left(\frac{k_{\mathrm{p}} 
    \eta_{\mathrm{s}}^2}{k_{\mathrm{f}} \sigma^{\ast} h} 
    + \dot{\Gamma}\frac{d\dot{\Gamma}}{d\Sigma_{\mathrm{E}}} \right),
\end{equation}
 the system has all bifurcation points at $\mu=\mu_{\mathrm{c}}=0$, with stable solutions for $\mu<0$ and periodic solutions for $\mu>0$.
\begin{figure*}
\centering
{\includegraphics[width=0.9\textwidth]{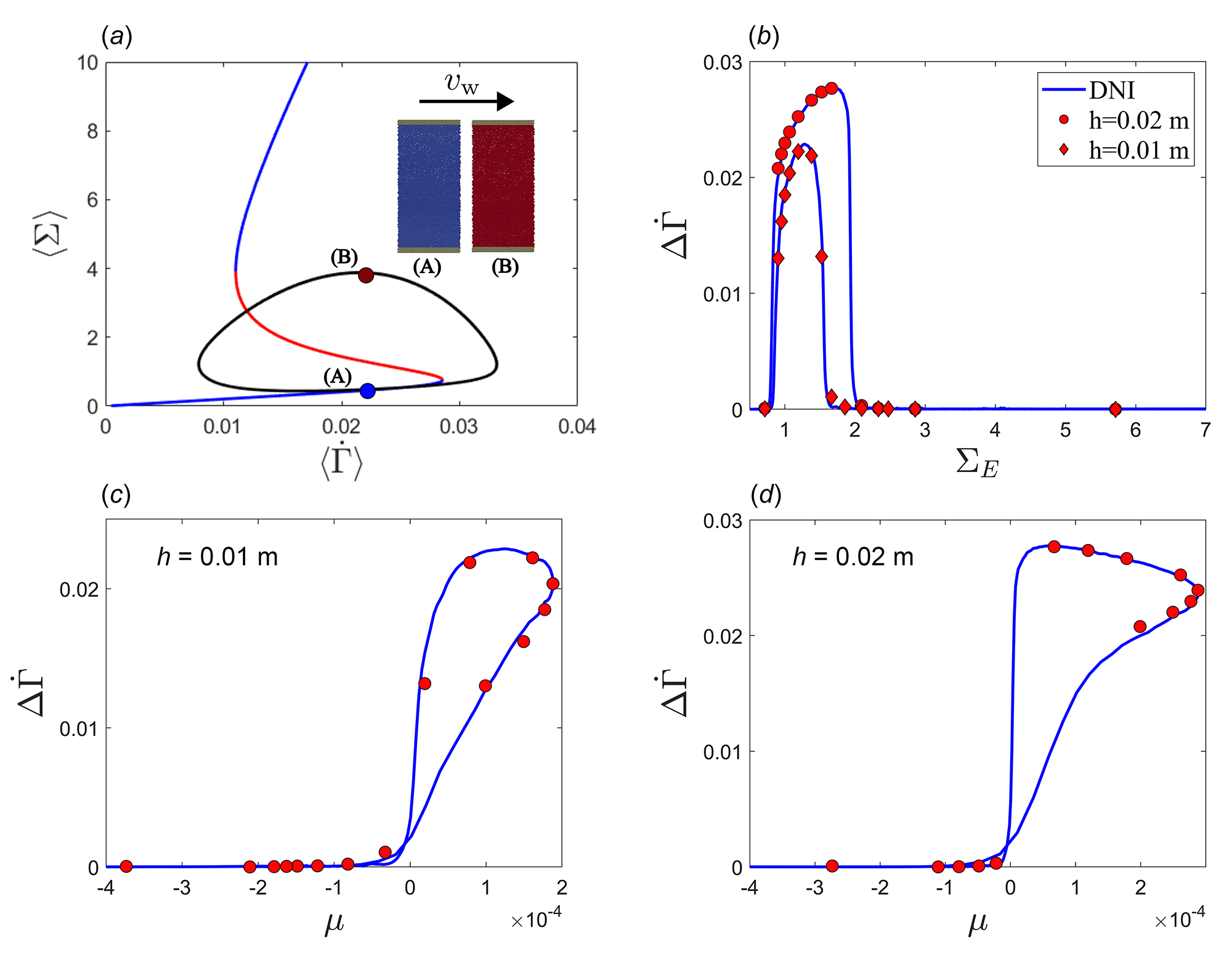}}
\caption{(a) A typical trajectory of an unstable solution exhibits periodic cycling between entirely frictionless and frictional domains. The trajectory is based on the spatially averaged stress and shear rate, and is plotted over a WC model flow curve where blue and red segments indicate regions of positive and negative gradients. The inset shows simulation frames of typical microstructural states $f$ during the minima (A) and maxima (B) of stress along the trajectory, where (A) is uniformly frictionless (blue) and (B) is uniformly frictional (red). (b) Comparison of limit cycle size $\Delta \dot{\Gamma}$ as a function of $\Sigma_{\mathrm{E}}$ along the range of the flow curve shows an increase in peak and broadening of the bifurcation points with increasing gap height. (c) and (d) show stability diagrams $\Delta \dot{\Gamma} (\mu)$ for $h=\SI{0.01}{\metre}$  $h=\SI{0.02}{\metre}$ respectively. Blue lines indicate solutions obtained by direct numerical integration, and red points are measurements from SPH simulations.
}
\label{fig1}
\end{figure*}
The stable solutions consist of an inward spiral towards the exact solution on the WC model, resulting in a steady state where the system lies exactly on the flow curve for any stress applied, regardless of the presence of DST \citep{Angerman2024}.
The bifurcation occurs when the response of the stress control mechanism is no longer fast enough to deal with the overshoots in the shear rate, leading to an increasing shear rate at the point of a large DST viscosity jump, which further increases the stress and thus drives the particles into an even more frictional state, initiating a positive feedback cycle of increasing viscosity and stress. Eventually, the direction of stress control will invert, initiating a similar feedback loop in the opposite direction, and completing the limit cycle as shown in Fig.\,\ref{fig1}(a).
To evaluate the capability of the SPH model to handle instabilities, we varied $k_{\mathrm{p}}$ to obtain unstable solutions and measured the amplitude of the apparent oscillations 
(Fig.\,\ref{fig1}(b)) in spatially-averaged dimensionless shear rate, $\Delta\dot{\Gamma}$, in both our SPH results and the results obtained via alternative direct numerical integration (DNI) of the ODEs (Eq.\,\eqref{eq: fdim evolution} and Eq.\,\eqref{ss dim}). 
Direct integration was carried out with the ODE45 Matlab algorithm with a relative tolerance of $\num{1e-10}$ and smoothed with the Savitzky-Golay algorithm with a window size of 5 subsequent points (less than 0.5\% of points) to correct the random numerical errors introduced by the solver.
Comparison between SPH simulations and the direct integration of the dynamics equations \eqref{eq: fdim evolution} and \eqref{ss dim} shows clear congruence between the two. The plots of the shear rate amplitude measured in SPH for both gap heights (Fig.\,\ref{fig1}(b--d)) show clear bifurcation in line with the predicted bifurcation condition, marked by a sudden increase in amplitude at $\mu=0$. 
Furthermore, the SPH model is capable of accurate predictions of exact limit cycle sizes (Fig.\,\ref{fig1}(b)) along the shear rate dimension for a range of imposed stress values ($\Sigma_{\mathrm{E}}$) spanning the entire frictionless-to-frictional transition of the flow curve. This agreement is illustrated in Fig. \ref{fig1}(b) for two different geometry gap heights.

\subsection{Vanishing Non-locality}
\label{weak non-local}

Having validated our implementation in the regime of homogeneous state, we can now consider instabilities with spatial heterogeneity. 
Here we consider the limit of vanishing non-locality.
Strictly, we set $l_{\mathrm{c}}=0$ and thus exclude non-locality entirely, but as we will show later, we can expect the following analysis to hold for $l_{\mathrm{c}} < L$, as any patterns with characteristic length below the resolution scale will not be resolved. 
In this and the following sections, we consider the regime of $\mu<0$ to investigate primarily spatial effects while excluding the temporal instability discussed in the previous section.

\begin{figure*}
\centering
{\includegraphics[width=0.9\textwidth]{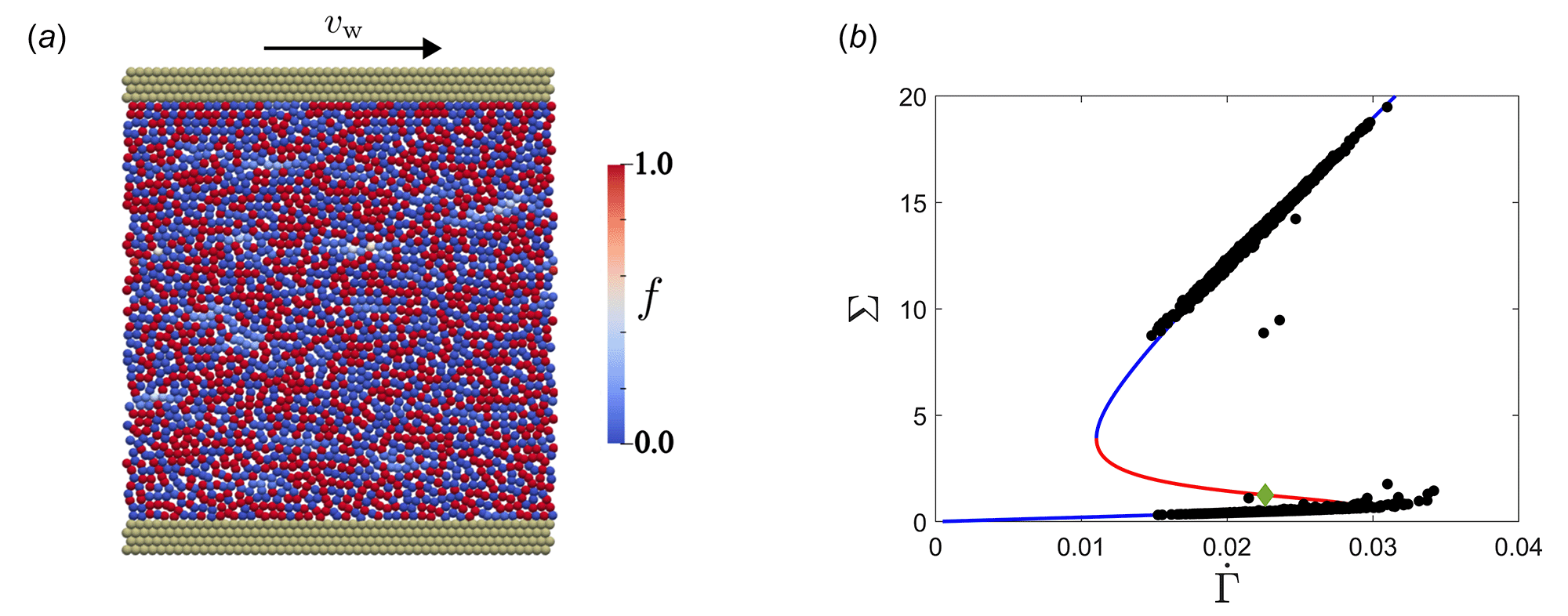}}

\caption{(a) The field visualisation of $f$ in steady state in shear rate controlled mode ~exhibits nondifferentiable microstructure (and stress) fields. 
(b) Individual SPH particles plotted on top of the theoretical WC curve. The particles split onto the two stable segments. Green diamond indicates initialisation state.}
\label{fig2}
\end{figure*}

We start by considering two particles in $\dot{\gamma}$--$\sigma$ phase space with coordinates $A(\dot{\gamma}^\ast,\sigma^\ast)$ and $B(\dot{\gamma}^\ast+\delta\dot{\gamma},\sigma^\ast)$, respectively, where particle $A$ lies exactly on the flow curve predicted by the WC model for a given input stress set-point (indicated by the superscript $\sigma^\ast$) and the particle $B$ receives a small positive perturbation in shear rate. 
Since the perturbation is horizontal, point $B$ preserves the stress such that
\begin{equation}
    \sigma_A=\sigma_B,
\end{equation}
which can be rewritten as
\begin{equation}
    \eta_A\dot{\gamma}^\ast=\eta_B(\dot{\gamma}^\ast+\delta\dot{\gamma}).
\end{equation}
This implies that the viscosity at $B$ must be less than the viscosity at $A$, from which we can infer that 
\begin{equation}
    \hat{f}-f_B>0
    \label{eq splitting}
\end{equation}
for any particle $B$ to the right of the flow curve. $\hat{f}$ is the steady state WC value at a given stress $\Sigma$, and $f^\ast$ indicates the steady state value corresponding to the applied stress $\Sigma_E$.  Notice that this is part of the local term in the microstructure evolution equation; hence, we can deduce that for any point to the right of the flow curve, the vector fields of the microstructure evolution must point directly upward in phase space, as the evolution of $f$ is positively correlated with the evolution of stress. 
By the same argument, we can also show that the vector fields to the left of the flow curve must always point downwards.
For flow curves with monotonous positive gradient 
($ d\dot{\gamma}/d\sigma >0)$, the resulting vector fields will always return any perturbed points to the flow curve. 
However, whenever negative gradients of the flow curve are present ($d\dot{\gamma}/d\sigma <0$), the vector field acting on the point will move it away from the flow curve. 
On the basis of this analysis, we can expect a spatial instability unique to the DST regime, where individual SPH particles receiving random numerical perturbations will `split' along the stress axis.

To confirm this behaviour, we performed rate-controlled simulations, in which the shear rate was maintained by a constant upper wall velocity. All particles were initialised with the same state exactly on the WC flow curve in the unstable region
The resulting field of $f$ (Fig.\,\ref{fig2}(a)) becomes discontinuous at the resolution level, with a lack of correlation in space between frictional and frictionless particles. 
In the $\dot{\gamma}$--$\sigma$ space, individual particles (Fig.\,\ref{fig2}(b)) are attracted to the stable Newtonian branches of the flow curve, where the vector fields act as a restoring force. 
The number of particles that split ``upwards'' and ``downwards'' is approximately equal because of the randomness in the numerical noise that places particles arbitrarily on either side of the WC flow curve.

\begin{figure*}
\centering
{\includegraphics[width=0.9\textwidth]{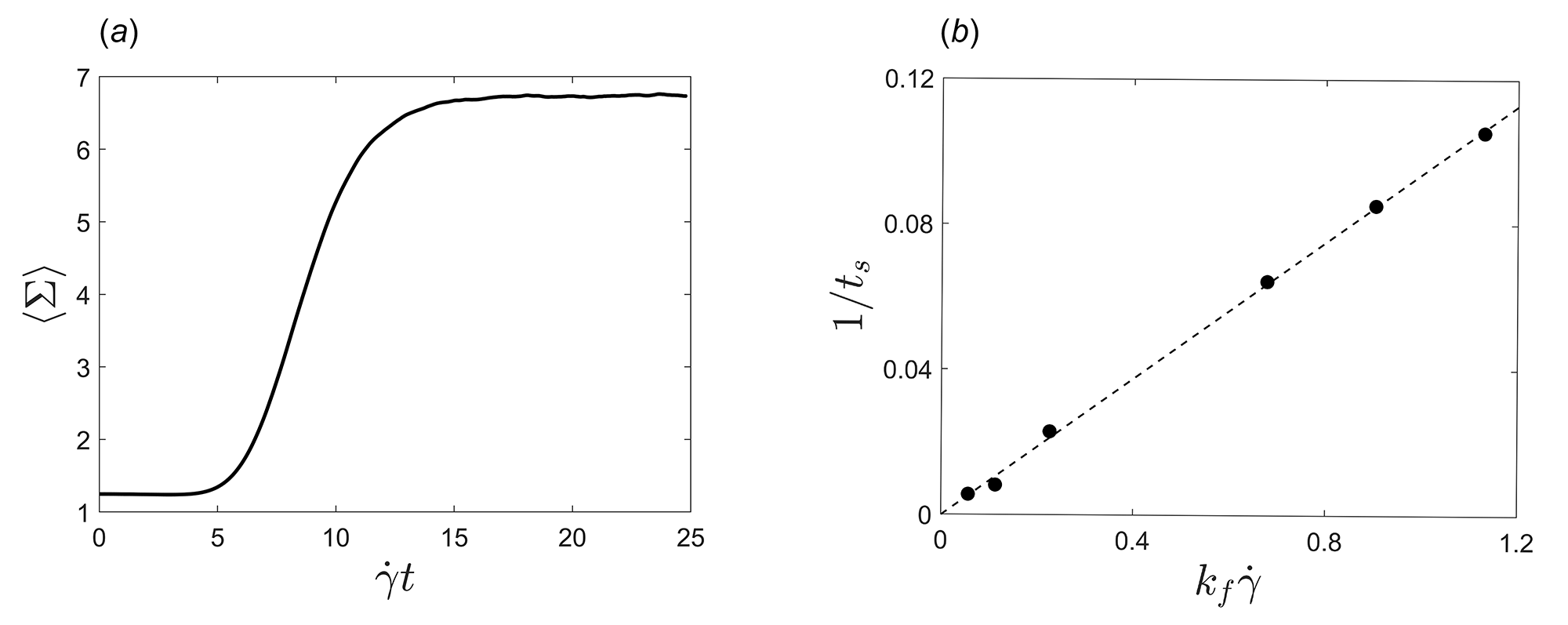}}

\caption{(a) Transient in average stress induced by the splitting of SPH particles for $k_{\mathrm{f}}=1$. (b) Plot of the rate of stress equilibration against the prefactor $k_{\mathrm{f}}\dot{\gamma}$ of local contributions to the microstructure evolution.}
\label{fig3}
\end{figure*}

The dynamics of the instability were probed by varying $k_{\mathrm{f}}$ in Eq.~\eqref{eq: f evolution}. 
The typical transient response during the instability (Fig.\,\ref{fig3}(a)) consists of a significant increase in the average stress within the domain, the lower stress plateau indicating the homogeneous initial state and the upper stress corresponding to the entirely split state.
The increase in stress is a result of an even number split in SPH particles along an asymmetric flow curve, with a much higher jump upward toward the frictional branch than the downward jump to the frictionless branch. 
In this case, the dynamics of stress evolution can be used as a proxy for the dynamics of the instability. 
For each value of $k_{\mathrm{f}}$, we define and measure a splitting equilibration time, $t_{\mathrm{s}}$, which corresponds to the time required to attain \SI{95}{\percent} of the upper stress value. 
The plot of inverse equilibration time (or the rate of equilibration) against the prefactor of the local term in the evolution (Fig.\,\ref{fig3}(b)) reveals a direct relation between the dynamics of stress-splitting and the dynamics of the local microstructure evolution, in line with the argument above based on Eq.\eqref{eq splitting}

\subsection{Moderate Non-Locality}
\label{moderate non-local}
\begin{figure*}
\centering
{\includegraphics[width=0.7\textwidth]{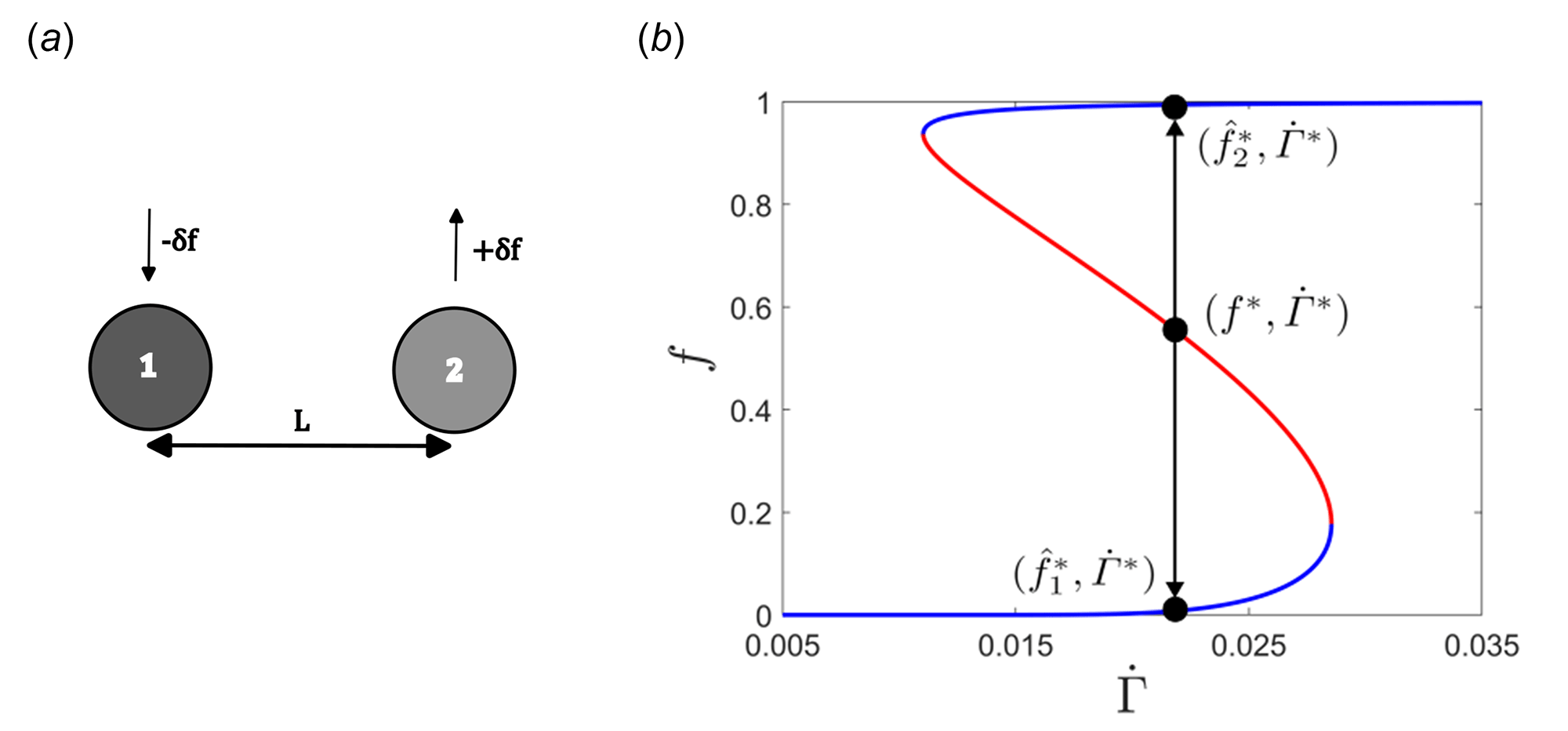}}
\caption{(a) Diagram of the one dimensional toy model used to conceptualise stress-splitting, and (b) diagram in $f$--$\dot{\Gamma}$ phase space of idealised stress-splitting motion.}
\label{fig4}
\end{figure*}

To evaluate the regime of moderate non-locality, where local microstructure growth and stress splitting is to some degree countered by diffusion, we begin by extending the simple two-particle model discussed above. We consider two SPH particles (Fig.\,\ref{fig4}(a)) $L$ apart, where $L$ is the resolution, both initialised on the flow curve with coordinates ($f^{\ast},\dot{\Gamma}^{\ast}$). Particle 1 receives a negative perturbation and begins a downward movement ($f^{\ast},\dot{\Gamma}^{\ast})\rightarrow (\hat{f}^{\ast}_1,\dot{\Gamma}^{\ast}$), while particle 2 receives a positive perturbation and begins an upward movement ($f^{\ast},\dot{\Gamma}^{\ast})\rightarrow (\hat{f}^{\ast}_2,\dot{\Gamma}^{\ast}$). 
This naive picture assumes that the shear rate is conserved (the split is purely vertical), which is not entirely representative of the instability as observed in Fig.\,\ref{fig2}(b), but allows us to simplify the problem. 
We also assume that the splitting proceeds in a way such that it can be captured with a single splitting parameter, $\beta$, that measures the extent of the split, allowing us to express the microstructure of each particle as interpolation between the homogeneous state ($\beta=0$) and the fully split state ($\beta=1$),
\begin{gather}
    f_1=(1-\beta)f^{\ast}+\beta \hat{f}_1,\\
    f_2=(1-\beta)f^{\ast}+\beta \hat{f}_2.
    \label{naive split eq}
\end{gather}
The steady state of stress-splitting can be expressed as a balance between the rate of splitting $(R_{\mathrm{split}})$ and the rate of microstructure diffusion ($R_{\mathrm{diff}})$ along the gradient created by splitting
\begin{equation}
    R_{\mathrm{split}} = R_{\mathrm{diff}}.
    \label{steady state eq}
\end{equation}
The rate of splitting can be written as
\begin{equation}
    R_{\mathrm{split}} = -k_{\mathrm{f}}
    \dot{\gamma}_1
    (\hat{f}_1-f_1)+k_{\mathrm{f}}
    \dot{\gamma}_2(\hat{f}_2-f_2),
    \label{split eq}
\end{equation}
where the first term represents `downward' movement of particle 1 and the second term indicates the `upward' movement of particle 2. 
We also know that $\dot{\gamma}_1=\dot{\gamma}_2$ since the split occurs along the lines of a constant shear rate. 
The diffusive term is the non-local part of the microstructure evolution equation.
\begin{equation}
    R_{\mathrm{diff}} = \alpha\nabla^2f
    \label{diff eq}
\end{equation}
After approximating the diffusive term using central difference method, and combining steady-state equations \eqref{steady state eq}, \eqref{split eq} and \eqref{diff eq}, along with the model of naive splitting \eqref{naive split eq}, we obtain
\begin{equation}
    L=\sqrt{\frac{\alpha}{k_{\mathrm{f}} \dot{\gamma}}}
    \sqrt{\frac{2\beta}{1-\beta}}.
    \label{master eq}
\end{equation}
The resulting equation \eqref{master eq} dictates the relation between the numerical resolution of the simulations, the characteristic length based on the relative magnitudes of local and non-local dynamics, and the resulting steady-state field smoothness captured by the splitting parameter $\beta$. 
This expression allows us to calculate the theoretical value of stress splitting based on the known microstructural dynamics.
\begin{equation}
    \beta=\frac{L^2}{L^2+l_{\mathrm{c}}^2}=\frac{L^2}{L^2+\frac{2\alpha}{k_{\mathrm{f}} \dot{\gamma}}}.
    \label{master beta eq}
\end{equation}

\begin{figure}
    \centering
    \includegraphics[width=0.5\textwidth]{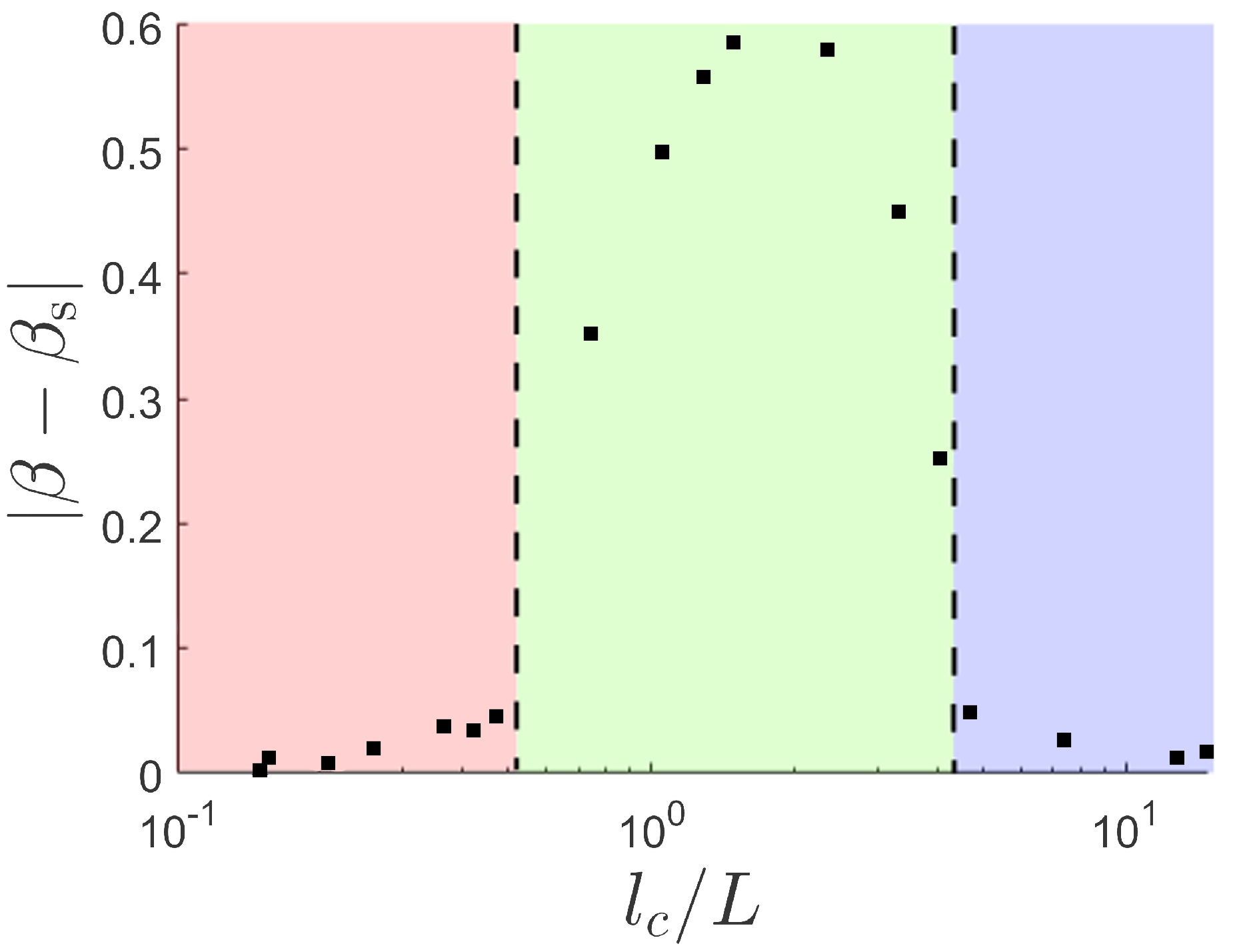}
    \caption{Plot of the difference between the model prediction of the field smoothness $\beta$ (Eq.\ref{master beta eq}) and the value measured in SPH simulations $\beta_{\mathrm{s}}$ (Eq.\ref{BETA SPH}) for a range of normalised characteristic length scales. The length scale was varied by changing the non-local parameter $\alpha$. Three distinct regions are highlighted based on the spatial pattern observed in the simulations: entirely discontinuous fields (red), continuous spatial patterns (green), and entirely homogeneous fields (blue).}
    \label{fig5}
\end{figure}
We can get an intuitive sense for what the equation \eqref{master eq} means: 
Consider a purely local case --- for vanishing  $\alpha$, $\beta$ trends towards $1$, which indicates a complete split. 
On the other hand, if $\alpha$ is large, $\beta$ approaches 0, which indicates a homogeneous field of $f$. 
Increasing the local contributions, $k_{\mathrm{f}}\dot{\gamma}$, would have the opposite effects. 
We could also pull particles apart --- increase $L$ by lowering the resolution --- which would reduce gradients and favour the local effects. 
This suggests that in presence of local and non-local effects, solutions are specific to the chosen resolution and $\alpha$.


In simulations, we cannot measure $\beta$ directly, but we can define an effective stress-splitting parameter $\beta_S$:
\begin{equation}
    \beta_{\mathrm{s}}=\frac{f_2-f_1}{\hat{f}_2-\hat{f}_1},
    \label{BETA SPH}
\end{equation}
where $f_2$ and $f_1$ are the measured maximum and minimum values in the field at steady state in a simulation. $\hat{f}_1(\Sigma_{\mathrm{E}})$ and $\hat{f}_2(\Sigma_{\mathrm{E}})$ are the known functions of input stress defined by the shape of the flow curve. 

\begin{figure*}
\centering
\includegraphics[width=1\textwidth]{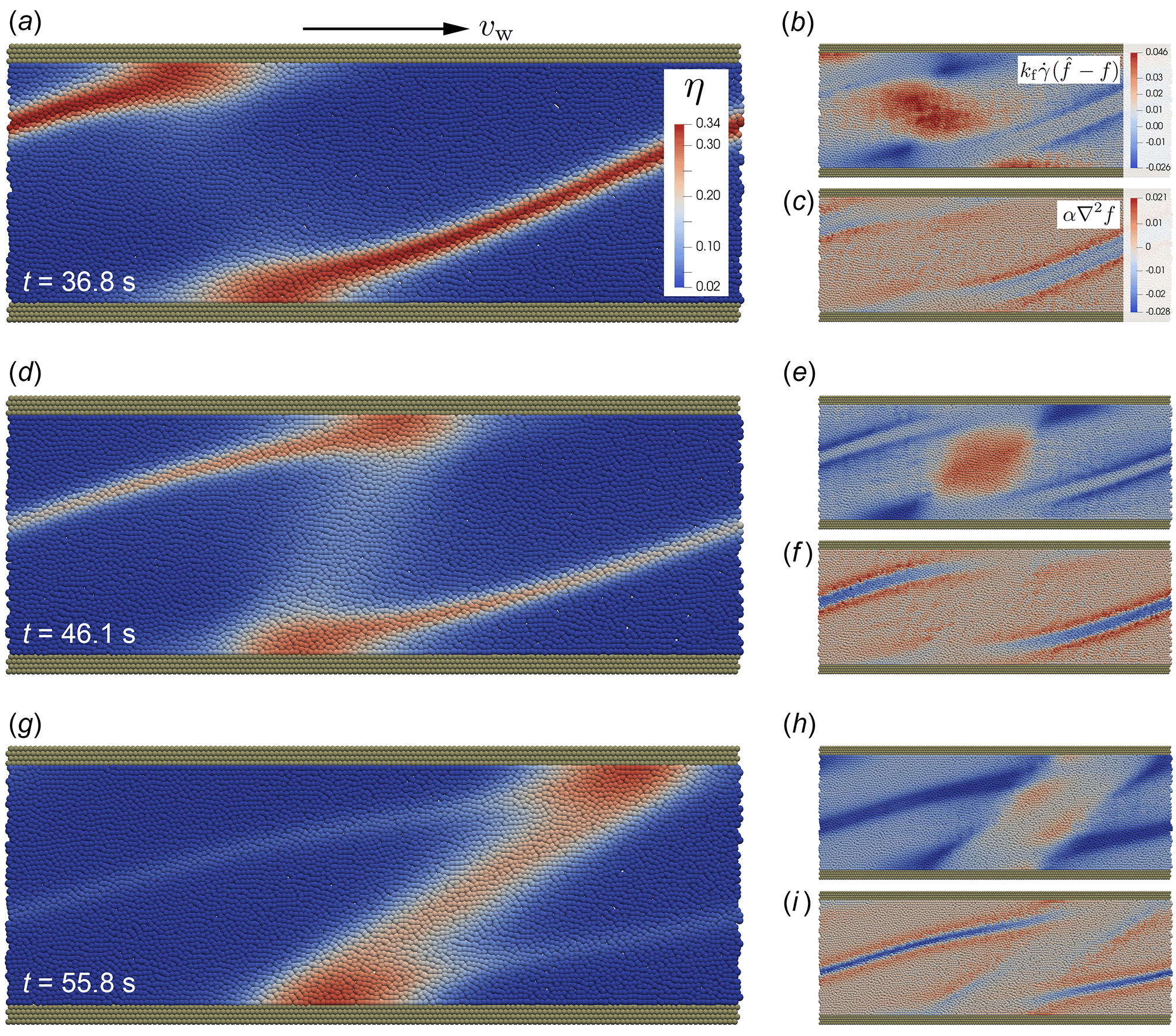}
\caption{Simulation time series showing SLP (red) at the top and bottom wall connected by a thin band. The large panels (a,d,f) show viscosity fields. The upper and lower smaller panels (b,c and e,f and h,i) show local and non-local contributions to microstructure formation as per Eq.~\ref{eq: f evolution} . The snapshots shown capture the process of a typical oscillation: just before constriction (a), during constriction (b), and after constriction (c). Simulation parameters: $l_{\mathrm{c}} \approx1.49L$ and $\Sigma_{\mathrm{E}}=1.25$.}
\label{fig6}
\end{figure*}
Stress-controlled simulations were carried out for a range of values of $\alpha$, at a constant stress set-point $\left(\Sigma_E=1.25\right)$. The degree of split in the resulting fields was measured by estimating the value of the stress-splitting parameter, $\beta_S$, through Eq.\eqref{BETA SPH}. The difference between the measured values in the simulations and the values predicted by the toy model, $\beta$, was plotted against the characteristic length scale, set in part by the varied $\alpha$ in Fig.\ref{fig5}. The plot shows good agreement in the extremes of dominant non-locality ($l_{\mathrm{c}}>5L$) and vanishing non-locality ($l_{\mathrm{c}} < 0.5L$).
Consider that the key assumption of the simple model is that all spatial effects can be resolved on the scale of a single particle pair --- the congruence between the model and simulations in these two extremes is expected as in both cases description on the level of resolution is sufficient to capture the overall picture of the field. 
In the intermediate region ($0.5L<l_{\mathrm{c}} < 5L$), the simulations drastically diverge from the model, from which we can infer that spatial correlation can no longer be described with only two particles --- extensive spatial patterns are formed while retaining inhomogeneous fields of microstructure.

The typical fields associated with this regime are visualised in Fig.\ref{fig6}, including fields of viscosity (Fig.\ref{fig6}(a,d,g)), local (Fig.\ref{fig6}(b,e,h)) and non-local (Fig.\ref{fig6}(c,f,i)) contributions to the rate of microstructure formation. The fields shown correspond to a simulation with typical characteristic length $l_c=1.49L$, and for a stable parameter configuration for the boundary-driven instability ($\mu=-0.0196$).
With the addition of moderate non-locality, the stress-splitting acts as a seeding instability, which after an induction period results in microstructure organised in clumps of high viscosity (or solid-like phases - SLPs) material attached to the wall boundaries. 
Both clumps travel in the direction of the flow with velocities approximately equal to the flow near their respective boundaries, resulting in a periodic scenario in which the upper SLP passes over the lower SLP, due to the periodic boundary in the flow direction, where the period of the passing is in part determined by the geometry length.  
Fig.\ref{fig6}(a) shows a situation where two previously formed SLPs on the upper and lower plate are approaching each other as the upper plate moves from left to right. The two SLPs are connected by a long band of high-viscosity (frictional state) material which formed in the previous cycle, the last time the SLPs passed each other. During the approach (Fig.\,\ref{fig6}(a)), SLPs begin forming a constriction thereby deforming the flow fields between them. 
This produces a region of elevated high-shear in the constriction space where the local term in the microstructure evolution equation (Fig.\ref{fig6}(b)) pushes $f$ towards the frictional branch, beginning the formation of a new high-viscosity connecting band as can be seen clearly in (Fig.\,\ref{fig6}(d)). 
Once the SLPs have passed each other and the constriction is resolved (Fig.\,\ref{fig6}(g)), the newly formed connecting band continues to increase in viscosity, while the previous old band is stretched thin allowing for diffusion to overcome the microstructure-building contributions localised within the band, and the band dissipates. 

\begin{figure*}
\centering
{\includegraphics[width=0.9\textwidth]{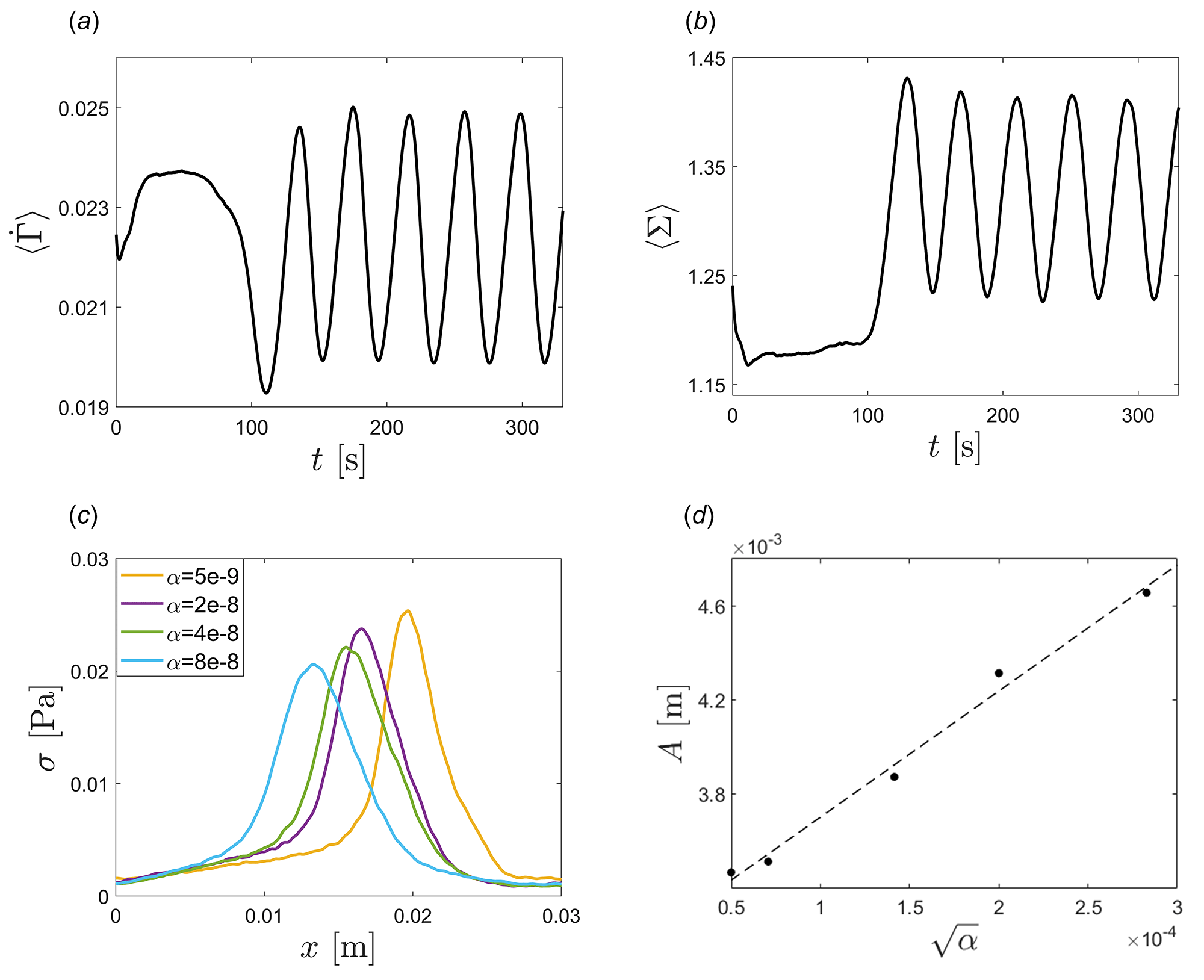}}

\caption{Typical average shear (a) and stress (b) measured during oscillations driven by relative motions of the SLPs, including the induction period of \SI{100}{\second}. 
(c) Spatial profiles of the upper SLP at varying strengths of non-locality. 
(d) Scaling of the SLP width on the strength of non-locality.}
\label{fig7}
\end{figure*}

During each pass, the average shear rate and stress oscillate (Fig.\,\ref{fig7}(a,b)) as a result of flow field deformation, and formation and dissipation of the connecting high-viscosity band. During each cycle, the shear rate between the time steps shown in Fig.\ref{fig6} (a, d, g) decreases, before again increasing. Looking at the initial field of local contributions to the microstructure formation (Fig.\ref{fig6}(b)), we observe that the formation of the connecting band begins during the approach and is driven by the local elements of the model. During the cycle, the viscous band appears in a subsequent time step, where the formation rate is not at its maximum (Fig.\ref{fig6}(e)), due to the delayed nature of microstructure formation. In general, as the shear rate decreases, the local contributions to the microstructure evolution decrease in magnitude (Fig.\ref{fig6} (b, e, h)), while the non-local effects are enhanced (Fig.\ref{fig6}(c,f,i)). This behaviour is in accordance with the results of the toy model, where the extent of splitting is mediated by the shear rate due to the varying balance of local and non-local forces. 

The present view of spatial patterns is broadly consistent with that proposed by \citep{Vikram2022} on the basis of boundary stress microscopy (BSM), wherein they infer the existence of SLPs at wall boundaries, connected by a region of high viscosity. 
The main difference here is that our band is aligned and stretched in the shear direction, rather than the compressive axis (although formation is initialised during compression). 
Their SLP formation can occur on both boundaries \citep{Vikram2017} or a single boundary, whereas in our simulations, with the particular parameter values described here, two SLPs --- one on each wall --- are always formed. 
Our model also does not display the rich dynamics of pattern formation in the wake of the SLP interaction \citep{Vikram2020}. 
Regardless, our simulations capture the broad picture of the spatial pattern and demonstrate a mode of temporal instabilities induced by the formation and evolution of a distinct spatial arrangement of the microstructure. 
Our result can also be viewed in congruence with the work of \citet{Nakanishi}, where similar geometry-spanning bands of microstructure were reported, and a mode of inhomogeneous oscillatory flow was presented.

The spatial pattern was further characterised by measuring the stress distribution along a line going through the SLP a small distance away from, and parallel to, the upper wall. (\SI{0.0005}{\metre}). 
As the non-locality ($\alpha$) is increased at constant local parameters, the SLPs become broader, while the peak stress inside the SLP decreases (Fig.\,\ref{fig7}(c)). 
Measured stress profiles were fitted to a normal distribution, measuring their standard deviation, $A$, as a proxy measurement for the characteristic length. 
The resulting fit (Fig.\,\ref{fig7}(d)) confirms SLPs grow in width with the square root of the diffusion coefficient,$\sqrt{\alpha}$, or in other words, linearly with the characteristic length scale Eq.\eqref{lc}.

A major difference between the present results and the experimental work on a stress-controlled system lies in the permanence of the SLPs. During BSM measurements in a \textit{shear rate} controlled system, the SLPs are persistent in time \citet{Vikram2020,Vikram2022}, but in a stress-controlled mode the SLPs are observed to dissipate during the oscillatory behaviour of the system. The proposed mechanism for this relies on the large stress fluctuations measured via BSM, which act through the stress-control scheme to sharply reduce the shear rate below the critical value, $\dot{\gamma}_c$, allowing SLPs to erode. The dissipation of SLPs will cause the total stress acting on the plate to diminish and the shear rate to increase, starting a new oscillatory cycle.

Our model should have all the component required to capture this proposed phenomenology. In figure \ref{fig3}(a) we demonstrated that the stress-splitting instability will act to increase the total average stress in the domain. A similar effect is expected to occur in the spatial pattern-forming regime: the appearance of SLPs is accompanied by a large increase in the average stress acting on the upper plate. The stress control will react to formation of SLPs by decreasing the shear rate, with the extent of shear rate reduction being proportional to the additional stress produced by the formation of the SLP. The response in field smoothness to the shear rate is characterised by Eq. \ref{master beta eq}. The decrease in shear rate will favour the non-local elements of the model, pushing the fields towards a homogeneous state and eroding the SLPs. The dissipation of SLPs will then in turn reduce total stress, potentially below the target stress set-point, resulting in an increase in shear rate. Increasing shear rate will drive the characteristic length scale from the homogeneous regime to a continuous spatial pattern regime (blue to green in Fig. \ref{fig5}) initiating a new cycle.

This phenomenology is not observed in Fig. \ref{fig6} likely due to the small amplitude of the oscillation in the shear rate, where a single cycle does not span a range of characteristic length scale wide enough to capture both the SLP-forming and homogeneous regimes. The underlying issue is the choice to artificially restrict the viscosity ratio in the rheology with the chosen parameters. In a real system, the viscosity ratio ($H=\eta(f=1)/\eta(f=0)$) defined as the ratio of the completely frictional state and the frictionless state is observed to span up to three orders of magnitude. In the simulations so far, we have chosen to limit the viscosity ratio to a moderate value ($H=31.77$) for numerical purposes. Extending this parameter to a more realistic value should result in higher 'excess' stress production as a result of SLP formation, and higher range of shear rates during oscillations. 

To test this, we simulated a new modified rheology by setting $\phi^m=10$. As such we can achieve a high viscosity ratio without exceedingly high values for the frictional viscosity, thereby avoiding excessive computational costs. The solvent viscosity is increased by 58\% to limit the maximum Reynold's number to less than 5. Dynamic parameters are adjusted as follows: $k_{\mathrm{f}}=20$, $\alpha=2\cdot10^{-7}$, and $k_{\mathrm{p}}=1$ (considered below as a "low-inertia" case). The new characteristic stress scale is reduced ($\sigma^{\ast}=5\cdot10^{-4}$) to maintain the ratio of inertial and microstructural timescales approximately constant ($\epsilon=0.025$). For a choice of input stress set-point of $\Sigma_E=10$, we obtain nearly equivalent characteristic length scale ($l_c/L=1.58$) to the simulations in Fig.\ref{fig6}, and we exclude the inertia-driven instability by setting $\mu=-0.00887$. The main differences in the new rheology are a significantly increased viscosity ratio ($H=583.51$) and a reduced diffusive timescale ($\omega=0.032$).
\begin{figure*}
\centering
{\includegraphics[width=0.9\textwidth]{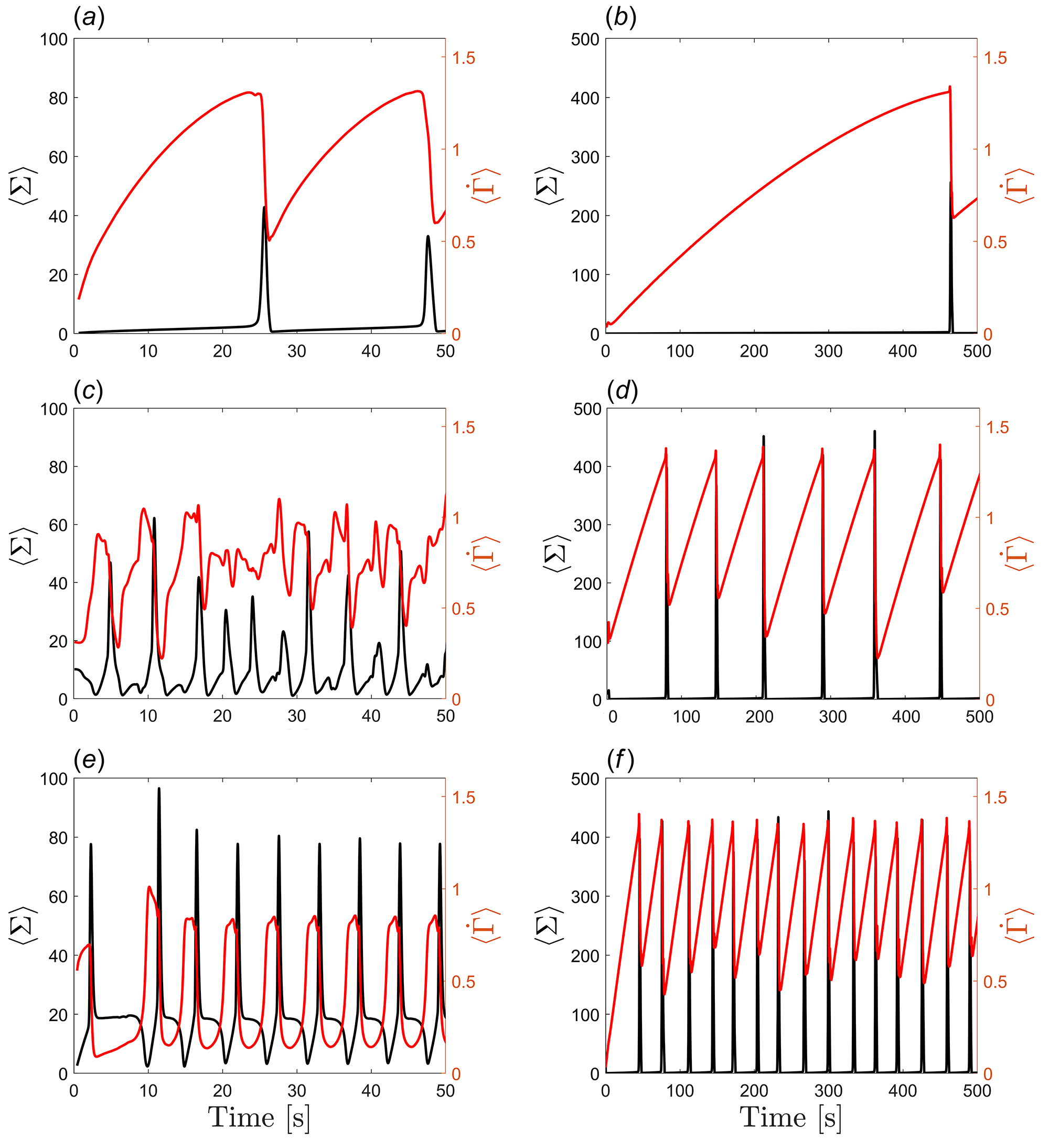}}
\caption{Normalised stress (black) and shear rate (red) signals with the modified rheology for (left column) $k_{\mathrm{p}}=1$ and (right column) $k_{\mathrm{p}}=0.01$. Solutions for three different stress set-point values shown for both cases: (a,b) $\Sigma_E=3$, (c,d) $\Sigma_E=10$, and (e,f) $\Sigma_E=20$.}
\label{fig8}
\end{figure*}

The results of this high viscosity ratio (high-$H$) simulation (Fig. \ref{fig8}) are in stark contrast to the previous low-$H$ simulations. Both shear rate and stress oscillate in an aperiodic high-frequency mode, with a significantly increased amplitude of both signals. Despite the chaotic nature of the signals, a clear trend can be observed in which an increase in the shear rate induces a sharp increase in stress, followed by a sudden drop in the shear rate. Similarly, the fields no longer display persistent, well-defined SLPs attached to the boundaries (as seen for the low-$H$ case in Fig. \ref{fig6}), but rather the microstructural patterns consist of geometry-spanning structures, which nucleate and dissipate accordingly with oscillations in shear and stress. The highly aperiodic nature of the oscillations in this case is a result of rapid increases in the shear rate during the SLP dissipation phase. As the plate accelerates, a thin layer of high shear rate forms near the upper wall resulting in a microstructure nucleation and propagation in pulses from the top to the bottom of the geometry. Comparing these results to the experimental dynamics in \citep{Vikram2020}, it is clear that the simulated dynamics do not display a clean distinction between a pattern-forming and homogeneous regimes such as observed experimentally. To more closely reproduce the experimental observations, we increase the inertia of the system by setting $k_{\mathrm{p}}=0.01$. Increasing the inertial timescale by two orders of magnitude should ensure ample time for microstructure dissipation. Note that now $\mu>0$ and we admit the possibility of a boundary-driven instability as in Section \ref{dominant non-local}.

\begin{figure*}
\centering
\includegraphics[width=1\textwidth]{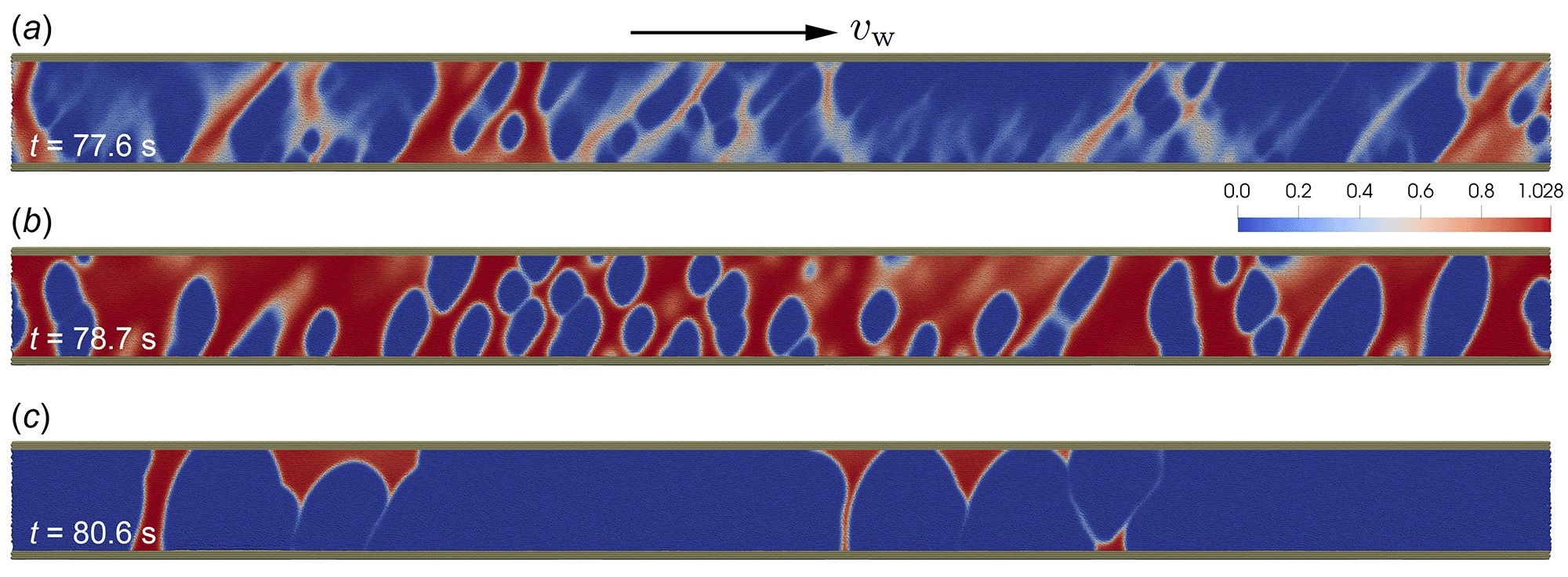}
\caption{Visualisation of the viscosity field for a modified high-viscosity ratio rheology with high system inertia ($k_p=0.01$). Three subsequent time steps of nucleation (a,b) and dissipation (c) of high-viscosity structures.}
\label{diss}
\end{figure*}

Figure~\ref{fig8} shows a comparison between low system inertia (a,c,d) and high inertia (b,d,f) stress and shear rate oscillations for three set-point stresses.In both high and low inertia cases, the signal retains significant aperiodic features and a comparable range of shear rate oscillations, but the pattern of increasing shear rate inducing a stress spike becomes even clearer for high inertia (Figure~\ref{fig8} (b,d,f)). Due to the slow action of the stress control scheme, the microstructure is allowed to form more extensive SLPs resulting in significantly higher and sharper increases in stress. In the same vein, oscillations in shear rate proceed by qualitatively regular (but quantitatively aperiodic) slow increases in shear, followed by sharp decreases in shear rate due to massive spikes in stress. This result shows remarkable congruence with the oscillatory patterns reported for a stress-controlled system (\citet{Vikram2020}), where the oscillations were accompanied by the formation and disappearance of high-stress events acting on the lower plate. Microstructural fields now feature clear separation between regimes, with the existence of spatial patterns confined to short periods of time associated with spikes in stress; periods of increasing shear rate (and low stress) are associated with entirely homogeneous fields of microstructure. Typical fields of viscosity during a microstructure nucleation and dissipation cycle are shown in Fig. \ref{diss} for a simple shear in a wide geometry with $l=15h$. No clear structures isolated to the boundaries can be observed. The microstructure spans the entire width of the gap with no clear alignment with either the extensional or compressive axes. Further simulations with an extended geometry ($l=15h$) were carried out to test the robustness of this phenomenology, with an similar patterns being observed, i.e. the observed patterns of microstructure displayed in Fig. \ref{diss} are independent of system width size. 

Varying the stress set-point ($\Sigma_E$) revealed a significant positive correlation between applied stress and the frequency of oscillations (Fig.\ref{fig8}(b,d,f)), a feature reported in the experimental work \citet{Vikram2020} - with no substantial alterations to the oscillation pattern in case of high inertia, but clear progression from near-periodic (Fig.\ref{fig8}(a)) to strongly chaotic (Fig.\ref{fig8}(c)) back to near-periodic solutions (Fig.\ref{fig8}(e)) in the case of low inertia.

\section{Conclusion}
We have presented the application of an SPH scalar DST continuum model to the problem of instabilities in simple shear. We checked the accuracy of the model using a previously reported instability mechanism and found excellent agreement with the relevant stability properties. Validation of our model at the extreme of dominant locality compared to the model of \citet{Richards} resulted in accurate predictions of both bifurcation points and limit cycle sizes in simulations, across a range of stress set points that span the entire flow curve, for two different gap height geometries. 

To overcome shortcomings of an entirely homogeneous view of the instability, we considered a spatial instability in the form of stress-splitting. We demonstrated that such an instability is unique for the DST regime and documented its origin in local contributions to the commonly used microstructural evolution equation. Although stress splitting reliably produced inhomogeneous flow fields in the region of negative flow-curve gradients, it also yielded nondifferentiable stress and microstructure fields. Inclusion of non-localityowed formation of continuous spatial patterns in microstructure. The form of non-locality in the present work was based on the formulation proposed by \citep{KamrinInclined}, but the analysis is expected to be general enough to anticipate alternate inclusions of non-locality. For example, particle migration as per suspension balance model should fulfil similar stress-diffusive role, with the added benefit of producing compaction fronts. With intermediate non-locality in the model, the obtained fields of microstructure showed formation of high viscosity phases near the solid geometry boundaries, connected by a band material in a highly frictional state. The simulated DST-rheology in this case was characterised by a relatively low viscosity contrast which severly limited the range of the amplitude of the shear rate oscillations. This produced long-lived SLPs, more akin to the expermentally observed SLPs for shear rate controlled systems \citep{Vikram2020}.

Simulations with high viscosity contrast (similar to realistic DST fluids) were able to capture the oscillatory phenomenology observed in the BSM measurements of \citet{Vikram2020}, along with commonly reported aperiodic solutions (\citet{Hermes2016}). Here, aperiodicity was found to be driven by the nucleation and dissipation of the microstructural patterns within the geometry, which were broadly similar but different in detail each cycle. The role of inertia was probed, and while it did not appear to have a determinitive role in the presence of instabilities in the range of parameters used in this work, it had a significant influence on the oscillatory patterns. It is noteworthy that this suggests that measurements of identical materials should have significantly different patterns based on the parameters such as the inertia of the used geometry or the rheometer itself. Across the range of input stresses in the entire multivariate region of the flow curve, high inertia simulations maintained their near-periodic (but still strictly aperiodic) oscillation patterns with significant increases in frequency as stress increased. In the case of low inertia, high-stress events increase in frequency, but now the oscillatory pattern transitions from near-periodic to chaotic and back to near-periodic oscillations with increasing stress.

The primary limitation of our implementation is the lack of vorticity direction. 
Much of the instabilities in DST have an aspect along the vorticity direction, both in experiments \citep{Herle,Vikram2022} and in simulations \citep{Fielding2018}. 
The inclusion of both gradient and vorticity instabilities is likely a key next step in future work toward a complete picture of rheochaos. 
In addition, the transition away from scalar models towards tensorial implementations could pose a significant advantage, allowing for distinguishing compressive and extensional shear effects in formation and destruction of the microstructure \citep{GW2019}.

\backsection[Funding]{PA and BS acknowledge funding from the Engineering and Physical Sciences Research Council [EP/S034587/1].
This research is partially supported by the Basque Government through the BERC 2022–2025 program and by the Spanish State Research Agency
through BCAM Severo Ochoa excellence accreditation CEX2021-0011
42-S/MICIN/AEI/10.13039/501100011033 and through the project
PID2020-117080RB-C55 (``Microscopic foundations of softmatter experiments: computational nano-hydrodynamics'' and acronym ``Compu-Nano-Hydro'').
R.S. acknowledges funding from the National Natural Science Foundation of China (12174390, 12150610463) and Wenzhou Institute, University of Chinese Academy of Sciences (WIUCASQD2020002).}

\backsection[Declaration of interests]{The authors report no conflict of interest.}

\backsection[Data availability statement]{The data and the simulation code that support the findings of this study are available from the corresponding author upon reasonable request.}

\bibliography{bib1}

\end{document}